\documentclass[a4paper,aps,amsmath,amssymb,nopacs,10pt]{revtex4}
\usepackage{epsfig,bm}
\usepackage{graphicx}
\usepackage{amsmath,amssymb,psfrag,textcomp}
\usepackage{amsthm,upgreek}

\newcommand{\bracket}[1]{\left\langle #1\right\rangle}

\newcommand{\be}{\begin{equation}}
\newcommand{\ee}{\end{equation}}
\newcommand{\bd}{\begin{displaymath}}
\newcommand{\ed}{\end{displaymath}}

\newcommand{\sgn}{{\rm sgn}}

\newcommand{\beeq}[1] {\begin{equation}\begin{split}#1\end{split}\end{equation}}
\newcommand{\beeqn}[1] {\begin{equation*}\begin{split}#1\end{split}\end{equation*}}
\begin{document}
\title{Reunion probabilities of $N$ one-dimensional random walkers with mixed boundary conditions}
\author{Isaac P\'erez Castillo}
\address{Department of Mathematics, King's College London, Strand, London WC2R 2LS, United Kingdom}
\address{Departamento de Sistemas Complejos, Instituto de F\'isica, UNAM, P.O. Box 20-364, 01000 M\'exico D.F., M\'exico}
\author{Thomas Dupic}
\address{Department of Mathematics, King's College London, Strand, London WC2R 2LS, United Kingdom}
\begin{abstract}
In this work we extend the results of the reunion probability of $N$ one-dimensional random walkers to include  mixed boundary conditions between their trajectories. The level of the mixture is controlled by a parameter $c$, which can be varied from $c=0$ (independent walkers) to $c\to\infty$ (vicious walkers). The expressions are derived by using Quantum Mechanics formalism (QMf) which allows us to map this problem into a  Lieb-Liniger gas (LLg) of $N$ one-dimensional particles. We use Bethe ansatz  and Gaudin's conjecture to obtain the normalized wave-functions and use this information to construct the propagator. As it is well-known,  depending on the boundary conditions imposed at the endpoints of a line segment, the statistics of the maximum heights of the reunited trajectories have some connections with different ensembles in Random Matrix Theory (RMT). Here we seek to extend those results and consider four models: absorbing, periodic, reflecting, and mixed. In all four cases, the probability that the maximum height is less or equal than $L$ takes the following form:
\begin{equation*}
F_N(L)=A_N\sum_{\bm{k}\in\Omega_{\text{B}}}\int D\bm{z} \exp\left[-\sum_{j=1}^Nk_j^2+\mathcal{G}_N(\bm{k})-\sum_{j,\ell=1}^N z_jV_{j\ell}(\bm{k})\overline{z}_\ell\right]
\end{equation*}
where $A_N$ is a normalization constant, $\mathcal{G}_N(\bm{k})$ and $V_{j\ell}(\bm{k})$ depend on the type of boundary condition, and $\Omega_{\text{B}}$ is the solution set of quasi-momenta $\bm{k}$ obeying the Bethe equations for that particular boundary condition.
\end{abstract}
\maketitle
\section{Introduction}
After more than a century since its introduction by Brown, Einstein, Langevin and Smoluchowski, the theory of random walks keeps finding an ever expanding range of applications. And yet ever so often one must be reminded that random walks are more than simply $\overline{x^2}=2Dt$, as there is more than meets the eye within the theory of Brownian motion. Sometimes there are hidden  mathematical gems waiting to be explored (e.g. linear probing with Hasting, queuing theory, applications to astrophysics \cite{Majumdar2005}). In other occasions we get beautiful mathematical surprises: who would have thought that the reunion probability of $N$ vicious walkers is proportional to the partition function of a two-dimensional Yang-Mills theory on a sphere with a given gauge group G, which is connected to the type of boundary condition on the vicious walkers' problem \cite{Forrester2011,Schehr2013}. Moreover, the resulting expression resembles those of the joint distribution of -discrete- eigenvalues in RMT, although connections between random walks and RMT are becoming anything but hardly a surprise \cite{Baik2000,Forrester2001,Johansson2003,Nagao2003,Katori2004,Ferrari2005,Tracy2007,Daems2007,Schehr2008,Novak2009,Nadal2009,Rambeau2010,Borodin2010,Bleher2011,Adler2012}.\\
In the present work we would like to extend the work done in \cite{Schehr2013}. We will consider the reunion probability of $N$ random walkers with mixed boundary conditions between trajectories. As we will see the boundary conditions are mathematically implemented so that by tuning one parameter we can go from the case of vicious walkers to the one of independent walkers.\\
The present paper is organized as follows: in sect. \ref{MD} we provide the definitions of the problem at hand. We briefly discussed the mapping to a LLg model in a box $[0,L]$ and discussed the various boundary conditions at then endpoints of this box. In sect. \ref{HB} we summarize the results obtained using Bethe ansatz and Gaudin's work to obtain normalized Bethe wave-functions. Finally in sect. \ref{RP} we use all these results to obtain a unified expression for the reunion probability. In the last section, we discuss future lines of research. 
\section{Model definitions}
\label{MD}
Let $\bm{x}(\tau)=(x_1(\tau),\ldots,x_N(\tau))$ be the vector of trajectories of $N$ one-dimensional random walkers. From the set of all possible trajectories, we are interested in the subset $\mathcal{B}$ which: (i) is restricted to the non-negative real line; (ii) obey certain boundary conditions between touching trajectories -to be defined below-; and (iii) all trajectories start together -or close enough to avoid mathematical complicacies- at $x=0$ at time $\tau=0$ and after exploring the space they reunite at $x=0$ -or again close enough to avoid unnecessary divergences- after a certain time which, without lose of generality, we set up to be $\tau=1$.\\
Within the subset $\mathcal{B}$ we focus on the probability that the maximum height of each trajectory is less or equal than $L$. Being mathematically more precise: for each random walk we define the random variable maximum height $H_{i}=\max_{0\leq \tau\leq 1}[x_i(\tau)]$ for $i=1,\ldots,N$ and we ask what is $\mathbb{P}[H_1,H_2,\ldots H_N\leq L]$ in the subset $\mathcal{B}$.  Using QMf this probability can we written as follows:
\beeq{
F_N(L)\equiv \mathbb{P}[H_1,H_2,\ldots H_N\leq L]=\lim_{\bm{\epsilon}\to\bm{0}}\frac{G_{[0,L]}(\bm{\epsilon},1|\bm{\epsilon},0)}{G_{\mathbb{R}^+}(\bm{\epsilon},1|\bm{\epsilon},0)}
\label{eq:RP}
}
where $\bm{\epsilon}=(\epsilon_1,\ldots,\epsilon_N)$ and where $G_{\mathcal{R}}(\bm{x},\tau|\bm{x_0},\tau_0)=\bracket{\bm{x}|e^{-H_{\mathcal{R}}(\tau-\tau_0)}|\bm{x_0}}$ is the propagator and $H_{\mathcal{R}}$ is the Hamiltonian of a one-dimensional LLg of $N$ particles:
\beeqn{
H_{\mathcal{R}}=-\sum_{i=1}^N\frac{\partial^2}{\partial x_i^2}+2c\sum_{i<j}\delta(x_i-x_j)+V_{\mathcal{R}}(\bm{x})
}
Here $V_{\mathcal{R}}(\bm{x})$ is a potential which confines the system into the region $\mathcal{R}$. In the restricted sector $0\leq x_1\leq x_2\leq \cdots\leq x_N\leq L$, the LLg Hamiltonian is that of free particles whose wave-function $\Psi(\bm{x})$ obeys the following set of mixed boundary conditions
\beeq{
\left(\frac{\partial}{\partial x_{j+1}}-\frac{\partial}{\partial x_{j}}-c\right)\Psi(\bm{x})\Bigg|_{x_{j+1}=x_{j}}=0\,,\quad\quad j=1,\ldots, N-1
\label{eq:LLbc}
}
with $\bm{x}=(x_1,\ldots, x_N)$. Going back to the context of random walks, it is precisely this set of boundary conditions that allows trajectories to cross with a certain rate controlled by c. Thus, as we can readily see from \eqref{eq:LLbc},  for $c\to\infty$ trajectories are not allowed to cross and we expect, as we will see, to recover the case for vicious walkers, while for $c=0$, the trajectories reflect, which is equivalent to independent random walkers.\\
\subsection{Boundary conditions and the endpoints of the box $[0,L]$}
As it has been nicely noted in previous works (see, for instance, the detailed analysis and discussions in \cite{Schehr2013}), there is a connection between vicious walkers and two-dimensional Yang-Mills theory on a sphere with a given gauge group, which depends on the boundary conditions imposed on the vicious walkers. Moreover, $F_N(L)$  mathematically resembles the expression of a joint distribution of -discrete- eigenvalues of a particular RMT ensemble. In this work, we do not try  to seek out further connections among these very interesting fields, but at the same time it would be foolish not to explore  how the boundary conditions in the line segment $[0,L]$ affects the reunion probability for the problem at hand.\\
With this in mind, and leaving for future work possible connections with other fields, we discuss four different  models. A note of caution here is in order: we use loosely the concept of reunion probability for the four models described below even though only for one it makes physical sense.\\
 Model I (absorbing case) corresponds to the  boundary conditions $\Psi(0,x_2,\ldots,x_N)=\Psi(x_1,\ldots,x_{N-1},L)=0$. Model II (periodic case) corresponds to take the periodic boundary conditions $\Psi(0,x_2,\ldots,x_N)=\Psi(x_2,\ldots,x_N,L)$.  Model III (reflecting case) corresponds to taking $\frac{\partial \Psi}{\partial x_1}\Big|_{x_1=0}=\frac{\partial \Psi}{\partial x_N}\Big|_{x_N=L}=0$ Finally, in model IV (mixed case) we consider mixed boundary conditions  $\left(\frac{\partial}{\partial x_1}-\alpha_1\right)\Psi\Big|_{x_1=0}=\left(\frac{\partial}{\partial x_N}-\alpha_N\right)\Psi\Big|_{x_N=L}=0$.\\
Before we proceed a few comments regarding these models are necessary. In the limit $c\to\infty$ (the so-called Tonks-Girardeau limit in the literature of integrable models) models I to III correspond precisely to models A to C in \cite{Schehr2013}. There is a slight difference, albeit inconsequential,  between model II and model B in \cite{Schehr2013}, which arises from the way periodicity is implemented in both models. For general values $c$, our results nicely extend those found in \cite{Schehr2013} and we are able to scan, by tuning the parameter $c$, the mathematical properties from the case of vicious walkers to independent ones.\\
Finally,  model IV is new. We have introduced this model as a mathematical toy in order to study the behavior of mixed boundary conditions  at the end points of the line segment $[0,L]$. Thus, even in the Tonk-Girardeau limit, the results of this model are new as they interpolate the effects  that the boundary conditions have between models A and C (or I and III for finite $c$).

\section{Hans Bethe's wave-functions and Gaudin's normalizing factors}
\label{HB}
To derive an expression for the Green function appearing in eq. \eqref{eq:RP} we follow the standard approach of expanding it by using a resolution of the identity in terms of normalized eigenfunctions of the Hamiltonian. For the LLg these are simply Bethe wave-functions \cite{Lieb1963,Takahashi2005} whose normalization factors can also be obtained, albeit in most cases these are simply conjectures, following the insightful work of Gaudin \cite{Gaudin1981}. After assuming that we obtain a complete basis and checking numerically that the expressions for the norms are most likely to be correct, we can then use these results to obtain expressions for the reunion probabilities. In what follows we summary the results for models from I to IV. The interested reader can find some hints to these derivations in the references provided. These and similar derivations can be found throughout the literature of the LLg (see, for instance, \cite{Gaudin1971,Korepin1982,Sklyanin1988,Batchelor2005,Nardis2013} and references therein)
\subsection{Model I. Absorbing case}
In this case the Bethe wave-function is given by
\beeq{
\Psi_{\text{I}}(\bm{x}|\bm{k})&=\frac{1}{\sqrt{\mathcal{N}_{\text{I}}}}\sum_{g}A_{\text{I}}(k_{g(1)},\ldots, k_{g(N)})e^{i\sum_{j=1}^N k_{g(j)}x_j}
\label{eq:modelIWF}
}
where $\sum_{g}$ represents the sum over the permutation and reflection group, $\mathcal{N}_{\text{I}}$ is the normalization factor while $A_{\text{I}}(k_{g(1)},\ldots, k_{g(N)})$ are the wave-function coefficients. After some tedious -but not arduous- derivation one arrives at the following expressions
\beeq{
A_{\text{I}}(k_{g(1)},\ldots,k_{g(N)})&=\sgn(g)\prod_{1\leq i<j\leq N}\left(k_{g(j)}-k_{g(i)}-ic\right)\left(k_{g(j)}+k_{g(i)}-ic\right)\\
\mathcal{N}_{\text{I}}&=C^{(\text{I})}_N\prod_{1\leq i<j\leq N}\left((k_i-k_j)^2+c^2\right)\left((k_i+k_j)^2+c^2\right)\\
&\det_{1\leq i,j\leq N}\left[\left(L+\frac{1}{2}\sum_{ i=1(i\neq j)}^{N}\left(D(k_j+k_i)+D(k_j-k_i)\right)\right)\delta_{j\ell}-\frac{1}{2}\overline{\delta}_{j\ell}\left(D(k_j-k_\ell)-D(k_j+k_\ell)\right)\right]
\label{eq:modelIAandN}
}
where $C^{(\text{I})}_N$ is a normalization constant, $\sgn(g)$ stands for the sign of the element $g$ of the permutation and the reflection, $D(x)=2c/(c^2+x^2)$,  $\det_{1\leq i,j\leq N}(A_{ij})$ stands for the determinant of the matrix $A$ with entries $A_{ij}$, and $\overline{\delta}_{j\ell}=1-\delta_{j\ell}$. Notice that we have preferred to used the notation $k_{g(i)}$ as opposed to $g(k_i)$ for purely aesthetic reasons. Note also that the value that $C^{(\text{I})}_N$ takes depends on which sector one is working on: if $1\leq x_{i}\leq L$ for $i=1,\ldots, N$ then $C^{(\text{I})}_N=2^N N!$; of course, in this case one must remember to symmetrize the wave-function \eqref{eq:modelIWF}, as this refers only to the restricted sector.\\ 
Finally, the vector of quasi-momenta $\bm{k}=(k_1,\ldots,k_N)$ obeys the Bethe equations.
\beeqn{
e^{2ik_{j}L}=\prod_{ \ell=1(\ell\neq j)}^{N}\frac{\left(k_{j}+k_{\ell}+ic\right)\left(k_{j}-k_{\ell}+ic\right)}{\left(k_{j}-k_{\ell}-ic\right)\left(k_{j}+k_{\ell}-ic\right)}\,,\quad\quad j=1,\ldots,N
}
It is instructive to discuss at this stage both, the Tonks-Girardeau  and the independent-walkers limits. In the first case (i.e. $c\to\infty$) it is quite straightforward to obtain that:
\beeq{
\Psi_{\text{I}}(\bm{x}|\bm{k})&=\frac{1}{\sqrt{N!}}\det_{1\leq i,j\leq N}\left[\sqrt{\frac{ 2}{L}}\sin(k_i x_j)\right]
\label{eq:FF}
}
while the Bethe equations yield $k_{j}=\frac{\pi n_j}{L}$ for $j=1,\ldots, N$, as found in \cite{Schehr2013}. On the other side, for $c=0$, we have instead that the wave-function takes the following form
\beeqn{
\Psi_{\text{I}}(\bm{x}|\bm{k})&=\frac{1}{\sqrt{N!}}\underset{1\leq i,j\leq N}{\text{per}}\left[\sqrt{\frac{ 2}{L}}\sin(k_i x_j)\right]
}
where $\underset{1\leq i,j\leq N}{\text{per}} (A_{ij})$ stands for the permanent of the matrix $A$ with entries $A_{ij}$. In this case the Bethe equations also yield $k_{i}=\frac{\pi n_i}{L}$.\\
These results, particularly the latter, can be derived in various ways. One possibility is to notice that the Bethe wave-function can be written -and this seems to be always true- as an operator acting on a effective vacuum, a wave-function of free fermions, which is precisely the wave-function \eqref{eq:FF}. Thus in the limit $c\to\infty$ the operator does not change this effective vacuum, while for $c=0$ the operator is such that the determinant is traded off by a permanent. The latter is intuitively meaningful as $c=0$ corresponds to independent random walkers, which in the QMf must be understood as a system of $N$ symmetric particles. Another way to check this is by  eye inspection of eq. \eqref{eq:modelIAandN}: one simply notes that the term $(k_{P(j)}^2-k_{P(i)}^2)$ in the wave-function coefficients will cancel the sign of the permutation and will transform the determinant into a permanent, while the overall factor $(k_j^2-k_i^2)$ for all pairs will cancel with the one coming from the normalization.
\subsection{Model II. Periodic case}
For this model the Bethe wave-functions reads
\beeqn{
\Psi_{\text{II}}(\bm{x}|\bm{k})&=\frac{1}{\sqrt{\mathcal{N}_{\text{II}}}}\sum_{P}A_{\text{II}}(k_{P(1)},\ldots, k_{P(N)})e^{i\sum_{j=1}^N k_{P(j)}x_j}
}
where $\sum_{P}$ represents the sum over all possible permutations of $(1,2,\ldots,N)$ and where
\beeqn{
 A_{\text{II}}(k_{P(1)},\ldots k_{P(N)})&=\sgn(P)\prod_{1\leq i<j\leq N}(k_{P(j)}-k_{P(i)}-ic)\\
\mathcal{N}_{\text{II}}&=C^{(\text{II})}_N\left[\prod_{i\neq j} \left(k_i-k_j +ic\right)\right]\det_{1\leq j,\ell\leq N}\left[\delta_{j\ell}\left(L+\sum_{i=1}^ND(k_j-k_i)\right)-D(k_j-k_\ell)\right]
 }
Here $\sgn(P)$ is the sign of the permutation and $C^{(II)}_{N}$ is a normalization constant. The vector of  quasi-momenta  obeys the following Bethe equations
\beeq{
e^{ik_jL}=\prod_{i=1(\ell\neq j)}^N\frac{k_i-k_j-ic}{k_i-k_j+ic}\,,\quad\quad j=1,\ldots, N
\label{BEperiodic1}
}
Let us again spend a couple of lines discussing the two extreme values for $c$. In the limit of vicious walkers we first note that the Bethe equations become
\beeqn{
e^{ik_jL}=(-1)^{N-1}\,,\quad\quad j=1,\ldots, N
}
which implies that $k_j=\frac{\pi n_j}{L}$ with $n_j$ $j=1,\ldots, N$ are arbitrary odd numbers for $N$ even or even numbers for $N$ odd.  Besides, one can convince oneself that the wave-function takes the following form:
\beeqn{
\Psi_{\text{II}}(\bm{x}|\bm{k})&=\frac{1}{\sqrt{L^NN!}}\det_{1\leq i,j\leq N}[\exp(ik_{i}x_j)]\,,\quad k_j=\frac{\pi n_j}{L}\,,\quad  \text{ $n_j$ even (odd) integers for odd (even) $N$}
}
This is more or less what is found in \cite{Schehr2013} but with two provisos. The first one is that we should notice that in our case the values allowed values for $n_i$ depend on whether $N$ is even or odd, while in \cite{Schehr2013} there is no such distinction. This is due to the fact that in their case they impose periodicity for each individual walker, while here we ask for periodicity of the system as a whole. The second difference is that in \cite{Schehr2013}, the authors work in polar coordinates. \\
On the other side for $c=0$ we find again that the determinant becomes a permanent \footnote{Actually this was already pointed out by Gaudin in \cite{Gaudin1971}.} and we can simply write:
\beeqn{
\Psi_{\text{II}}(\bm{x}|\bm{k})&=\frac{1}{\sqrt{L^NN!}}\underset{1\leq i,j\leq N}{\text{per}}[\exp(ik_{i}x_j)]\,,\quad 
}
while the Bethe equations yield $k_j=\frac{2\pi n_j}{L}$, where $n_j$ can be any integer.
\subsection{Model III. Reflecting case}
The results for this model are obviously very similar to those of model I the only difference between in the sign that appears in the wave-function coefficients. These are:
\beeqn{
A_{\text{III}}(k_{g(1)},\ldots,k_{g(N)})&=\sgn(P)\prod_{1\leq i<j\leq N}\left(k_{g(j)}-k_{g(i)}-ic\right)\left(k_{g(j)}+k_{g(i)}-ic\right)
}
The rest -the expression for the norm and the Bethe equations- is the same as model I. As expected, the Tonks-Girardeau and the independent-walkers limits are as those of model I: one simply has to change the sine function by the cosine function.
\subsection{Model IV. Mixed case}
Finally, for  model IV we have that the Bethe wave-function takes the following form
\beeqn{
\Psi_{\text{IV}}(\bm{x}|\bm{k})&=\frac{1}{\sqrt{\mathcal{N}_{\text{IV}}}}\sum_{g}A_{\text{IV}}(k_{g(1)},\ldots, k_{g(N)})e^{i\sum_{j=1}^N k_{g(j)}x_j}
}
where the wave-function coefficients and the norm read:
\beeqn{
A_{\text{IV}}(k_{g(1)},\ldots,k_{g(N)})&=\sgn(g)\left[\prod_{1\leq i<j\leq N}(k_{g(j)}-k_{g(i)}-ic)(k_{g(j)}+k_{g(i)}-ic)\right]\prod_{j=1}^N(k_{g(j)}-i\alpha_1)\\
\mathcal{N}_{\text{IV}}&=C_{\text{IV}}^{(N)}\left[\prod_{1\leq i<j\leq N}\left((k_i-k_j)^2+c^2\right)\left((k_i+k_j)^2+c^2\right)\right]\prod_{j=1}^N(k^2_{j}+\alpha^2_1)\\
&\det_{1\leq i,j\leq N}\Bigg[\left(L+\frac{\alpha_1}{k^2_{j}+\alpha^2_1}-\frac{\alpha_N}{k^2_{j}+\alpha^2_N}+\frac{1}{2}\sum_{ i=1(i\neq j)}^{N}\left(D(k_j+k_i)+D(k_j-k_i)\right)\right)\delta_{j\ell}\\
&-\frac{1}{2}\overline{\delta}_{j\ell}\left(D(k_j-k_\ell)-D(k_j+k_\ell)\right)\Bigg]
}
Here the quasi-momenta obey the following set of Bethe equations
\beeqn{
e^{2ik_{j}L}&=\frac{(k_{j}+i\alpha_1)(k_{j}-i\alpha_N)}{(k_{j}-i\alpha_1)(k_{j}+i\alpha_N)}\prod_{\ell=1(\ell\neq j)}^N\frac{(k_{j}+k_{\ell}+ic)(k_{j}-k_{\ell}+ic)}{(k_{j}-k_{\ell}-ic)(k_{j}+k_{\ell}-ic)}\,,\quad\quad j=1,\ldots,N
}
Let us pause here for a second and perform again some consistency checks. We first note that in the limit $\alpha_1,\alpha_N\to\infty$  we indeed recover model I. On the other side, Model III  is recovered by taking the limit $\alpha_1=\alpha_N=0$ as, in this case,  the factor $\sgn(g)\prod_{j=1}^N k_{g(j)}$ in the wave function coefficient will give $\sgn(P)\prod_{j=1}^N |k_j|$, with the factor $\prod_{j=1}^N |k_j|$ canceling the one coming form the norm.\\
It is worth noticing that this model may bring news results even for vicious walkers. Indeed, in the limit $c\to\infty$ the wave-function can be expressed in the following operator form
\beeqn{
\Psi_{\text{IV}}(\bm{x}|\bm{k})&=\frac{1}{\sqrt{N!\prod_{j=1}^N(k^2_{j}+\alpha^2_1)\left[1+\frac{1}{L}\left(\frac{\alpha_1}{k^2_{j}+\alpha^2_1}-\frac{\alpha_N}{k^2_{j}+\alpha^2_N}\right)\right]}}\prod_{j=1}^N\left(\frac{\partial}{\partial x_j}+\alpha_1\right)\det_{1\leq i,j\leq N}\left[\sqrt{\frac{2}{L}}\sin(k_i x_j)\right]
}
with Bethe equations
\beeqn{
e^{2ik_{j}L}&=\frac{(k_{j}+i\alpha_1)(k_{j}-i\alpha_N)}{(k_{j}-i\alpha_1)(k_{j}+i\alpha_N)}\,,\quad\quad j=1,,\ldots,N
}
This latter case simplifies even further when we consider $\alpha_1=\alpha_N\equiv \alpha$, obtaining:
\beeqn{
\Psi_{\text{IV}}(\bm{x}|\bm{k})&=\frac{1}{\sqrt{N!\prod_{j=1}^N(k^2_{j}+\alpha^2)}}\prod_{j=1}^N\left(\frac{\partial}{\partial x_j}+\alpha\right)\det_{1\leq i,j\leq N}\left[\sqrt{\frac{2}{L}}\sin(k_i x_j)\right]
}
while from the Bethe equations we obtain $k_i=\frac{\pi n_i}{L}$. Yet these wave-functions are very different to the ones obtained in \cite{Schehr2013}. The obvious question is then how this affects the results for the reunion probabilities.
\section{Reunion probabilities for models from I to IV}
\label{RP}
We have all ingredients needed to obtain expressions for the reunion probabilities. The next step is to write the Green function as $G_{[0,L]}(\bm{\epsilon},1|\bm{\epsilon},0)=\sum_{\bm{k}\in\Omega^{(a)}_{\text{B}}}||\Psi_{a}(\bm{\epsilon}|\bm{k})||^2e^{-\sum_{j=1}^N k_j^2}$, where we have denoted as $\Omega^{(a)}_{\text{B}}$ the solution set of vectors of quasi-momenta obeying the Bethe equations of model $a\in\{$I, II, III, IV$\}$. To perform the limit $\bm{\epsilon}\to\bm{0}$, the easiest approach is to express  the Bethe wave-function in operator form acting on a effective vacuum. This vacuum will depend on the type of boundary condition but it will generally be expressed in terms of a Slater's determinant. Expanding this determinant in power series around zero allows us to untangle the position vector with the vector of quasi-momenta (Nb. actually one obtains the product of two determinants for positions and quasi-momenta whose powers are coupled by multiple sums). After some algebra, and after massaging the expressions for purely aesthetic and mathematically evoking reasons, one is able to write the reunion probability as follows:
\beeq{
F^{(a)}_N(L)=A^{(a)}_N\sum_{\bm{k}\in\Omega^{(a)}_{\text{B}}}\int D\bm{z} \exp\left[-\sum_{j=1}^Nk_j^2+\mathcal{G}^{(a)}_N(\bm{k})-\sum_{j,\ell=1}^N z_jV^{(a)}_{j\ell}(\bm{k})\overline{z}_\ell\right]\,,\quad\quad a\in\{\text{I, II, III, IV}\}
\label{eq:final}
}
with the following expressions for $\mathcal{I}^{(a)}_N(\bm{k})\equiv e^{\mathcal{G}^{(a)}_N(\bm{k})}$ and the two-body potential $V^{(a)}_{ij}(\bm{k})$: 
\beeqn{
\mathcal{I}^{(\text{II})}_{N}(\bm{k})&=\prod_{1\leq i<j\leq N}\frac{(k_i-k_j)^2}{(k_i-k_j)^2 +c^2}\,,\quad \mathcal{I}^{(\text{III})}_{N}(\bm{k})=\prod_{1\leq i<j\leq N}\frac{(k^2_i-k^2_j)^2}{[(k_i-k_j)^2 +c^2][(k_i+k_j)^2 +c^2]}\\
\mathcal{I}^{(\text{I})}_{N}(\bm{k})&=\mathcal{I}^{(\text{III})}_{N}(\bm{k})\prod_{i=1}^Nk_i^2\,,\quad \mathcal{I}^{(\text{IV})}_{N}(\bm{k})=\mathcal{I}^{(\text{III})}_{N}(\bm{k})\prod_{i=1}^N\frac{k_i^2}{k_i^2+\alpha_1^2}
}
and
\beeqn{
V^{(\text{I})}_{j\ell}(\bm{k})&=V^{(\text{III})}_{j\ell}(\bm{k})=\left(L+\frac{1}{2}\sum_{ i=1(i\neq j)}^{N}\left(D(k_j+k_i)+D(k_j-k_i)\right)\right)\delta_{j\ell}-\frac{1}{2}\overline{\delta}_{j\ell}\left(D(k_j-k_\ell)-D(k_j+k_\ell)\right)\\
V^{(\text{II})}_{j\ell}(\bm{k})&=\delta_{j\ell}\left(L+\sum_{i=1}^ND(k_j-k_i)\right)-D(k_j-k_\ell)\,,\quad V^{(\text{IV})}_{j\ell}(\bm{k})=\left(L+\frac{\alpha_1}{k^2_{j}+\alpha^2_1}-\frac{\alpha_N}{k^2_{j}+\alpha^2_N}\right)\delta_{j\ell}+V^{(\text{I})}_{j\ell}(\bm{k})
}
\section{Summary and future work}
\label{SF}
In this simple work we aimed to generalizing the expressions of reunion probabilities of $N$ vicious walkers by allowing the walkers the opportunity to cross. We have followed the standard QMf to map the problem into a LLg and, thanks to the godsend techniques of Bethe ansatz and the work of Gaudin, we have been able to obtain exact expression of $F_N(L)$ for finite $N$ and valid for any non-negative value of $c$. From a purely mathematical point of view we see how our new formulas nice interpolate between vicious walkers ($c=\infty$) and independent walkers ($c=0$).\\
Unfortunately there is still a long way to go and many questions need to be answered: first and foremost one needs to check whether the new  found formulas are correct. This could be done at least numerically, but one can foresee two problems, one minor the other one major. We first need to find a way to evaluate numerically  \eqref{eq:final}, which involves  solving the Bethe equations for each model and then evaluate numerically the sums and integrals in \eqref{eq:final}. This can be reasonably achieved. The problem seems to find an independent way to evaluate the reunion probability, so a comparison between the two results could be done. The obvious choice would be to simulate the set of $N$ random walks with mixed boundary conditions and then extract from there the statistics of the reunion probability. However, this is not an easy numerical task.\\
Secondly, and in the spirit of \cite{Schehr2013}, one would like to study the typical and large fluctuations for large $N$ of the reunion probability \eqref{eq:final} using, for instance, Coulomb gas approach. A simple eye inspection of \eqref{eq:final} suggests that this seems to be an easy task. However, one must remember that, unlike \cite{Schehr2013}, the vector of quasi-momenta must obey the Bethe equations. Thus, in the limit for large $N$( and large $L$) this must be taken into account. This, in turn, is likely to involve the density of bosons in some way.\\
Finally, and this is at the moment pure elucubration, we wonder whether there exists an extension of a two-dimensional Yang-Mills theory on sphere whose partition function is proportional to \eqref{eq:final}.\\
These research lines are currently being looked at.

\acknowledgements
IPC would like to thank Satya Majumdar for all the interesting and exciting discussions we had during his week-long stay at King's College London. He also would like to thank Jean-S\'ebastien Caux for his interest in this work and for pointing out some references. TD would like to thank his home university for allowing him to do an internship at KCL and also thanks the hospitality of the Disordered Systems Group at KCL.

\bibliographystyle{apsrev}
\bibliography{bib}

\end{document}